# Nanomechanical Detection of Itinerant Electron Spin Flip

Guiti Zolfagharkhani[1], Alexei Gaidarzhy[2], Pascal Degiovanni[1,3], Stefan Kettemann[4], Peter Fulde[5] and Pritiraj Mohanty[1]

[1]*Department of Physics, Boston University, 590 Commonwealth Avenue, Boston, MA 02215, USA*
[2]*Department of Mechanical and Aerospace Engineering, Boston University, 110 Cummington Street, Boston, MA 02215, USA*
[3]*Universite de Lyon, CNRS-Laboratoire de Physique de l'ENS de Lyon, Ecole Normale Sup´erieure de Lyon, 46, All´ee d'Italie, 6964 Lyon cedex 07, France*
[4] *Jacobs University Bremen, School of Engineering and Science, Campus Ring 1, D-28759 Bremen, Germany.*
[5]*Max-Planck Institut für Physik Komplexer Systeme, Nöthnitzer Str. 38, D-01187, Dresden, Germany*

**Spin is an intrinsically quantum property, characterized by angular momentum. A change in the spin state is equivalent to a change in the angular momentum or mechanical torque. This spin-induced torque has been invoked as the intrinsic mechanism in experiments[1,2] ranging from the measurements of angular momentum of photons[3], *g*-factor of metals[4,5,6,7] and magnetic resonance[8] to the magnetization reversal in magnetic multi-layers[9,10,11,12,13,14,15]. A spin-polarized current introduced into a nonmagnetic nanowire produces a torque associated with the itinerant electron spin flip. Here, we report direct measurement of this mechanical torque and itinerant electron spin polarization in an integrated nanoscale torsion oscillator, which could yield new information on the itinerancy of the *d*-band electrons. The unprecedented torque sensitivity of $10^{-22}$ N-m/$\sqrt{\text{Hz}}$ may enable applications for spintronics, precision measurements of CP-violating forces[16,17], untwisting of DNA[18] and torque generating molecules[19,20].**

Extensive studies of spin transfer and spin relaxation at a ferromagnetic-nonmagnetic interface[21,22,23] have shown that such a system can act as an effective source or sink of angular momentum in the presence of an electric current[24,25]. Consider a device involving a hybrid metallic nanowire whose left half is ferromagnetic (FM) and right half is nonmagnetic (NM) (Fig. 2a). Since the ferromagnet is magnetized by an axial magnetic field $B$ along the wire, this arrangement allows spin polarization in the ferromagnetic segment and spin flip in the nonmagnetic segment of the nanowire[26]. When a current $I$ is driven through the wire, non-equilibrium spin density accumulation $\delta m = \frac{\mu_B \Delta N_{noneq}}{V} = \mu_B \frac{I_s}{e}\frac{\tau_{sf}}{V}$ is produced near the FM-NM interface as a result of the balance between spin injection and spin relaxation processes[22]. Here $\mu_B$ is the Bohr magneton, $V$ is the volume, $\tau_{sf}$ is the spin relaxation time, and $\Delta N_{noneq}$ is the number of non-equilibrium spins. The spin-polarized injection current is $I_s = I_{up} - I_{down} = P_I \times I$, where $P_I = \frac{I_{up} - I_{down}}{I}$. Therefore, spin-flip transfer torque is given by[24,26]:

$$\vec{T}_{SF} = \frac{\Delta N_{noneq}}{2}\frac{\Delta L}{\Delta t}\hat{z} = \frac{\hbar}{2}\frac{I}{e}P_I \hat{z}, \qquad \text{(Eq.1)}$$

where $\Delta L$ is the angular momentum change.

Here, we demonstrate a nanomechanical device designed to detect and control spin-flip torque. Fig. 1a shows the scanning electron micrograph of a single-crystal silicon torsion oscillator, fabricated by electron-beam lithography and surface nanomachining. The FM-NM interface is located at the junction of the cobalt (Co) and gold (Au) electrodes on the central wire. When current is driven through the interface via electrical connections 3-4, the spin-flip process causes localized mechanical torque since the spin diffusion length is much smaller than the central wire length. The outer electrode 1-2 is used to detect the transverse displacement of the outer torsion element magnetomotively. We extract the spin torque from the magnetic field dependence of the induced voltage $V_{emf}$ on the outer electrode, which is amplified and measured by a lock-in amplifier (Fig. 1c). The torsion oscillator is mounted at the center of a 16-tesla superconducting solenoid magnet on a movable sample stage, capable of controllably tilting the sample in the plane of the oscillator. The tilt angle $\phi$ between the applied magnetic field and the $\hat{z}$ axis of the structure can then be varied from 0 to $90^\circ$ with a precision of $\pm 1^\circ$ as shown in Fig. 1c. The field component $B_\perp = B\sin(\phi)$ perpendicular to the detection electrode 1-2 is the effective field that induces the magnetomotive voltage $V_{emf}$. The magnetic field polarizes the central wire in the magnet leading to polarization $P_I = P\tanh\left(\frac{\chi\left(B \pm B_0\right)}{\mu_0 M}\right)$, where $P$, $\chi$, $B_0$, and $M$ are the saturation polarization, susceptibility, coercive field, and magnetization of the ferromagnet, respectively. Our measurement setup is sensitive to the polarization of the central wire along the wire axis $P_I(B) = P_I\cos\phi$. The sample stage and the coaxial cables are thermally anchored to the mixing chamber of a dilution cryostat in vacuum at a temperature of 110 mK. The vibration spectrum of the torsion oscillator shows two resonance peaks at 5.06 MHz and 6.71 MHz corresponding respectively to the symmetric and anti-symmetric torsion modes with typical quality factors $Q = 28000$.

Eq. 2 captures the dynamical response of the spin-torque resonator in the magnetomotive actuation-detection setup. It is derived by modeling the resonator as a damped harmonic oscillator with response contribution from the relevant modes as the oscillator is driven in the anti-symmetric resonance mode by an applied current (Supplementary section B). The induced Faraday voltage on resonance is given by

$$V(\omega_0) = \frac{-i\omega_0 LdB_\perp}{J}\left(\frac{-LdB_\perp\lambda_J}{\Omega_0^2-\omega_0^2+i\Gamma\omega_0}+\frac{\hbar P_I/2e}{i\gamma\omega_0}\right)I(\omega_0)\,. \qquad \text{(Eq. 2)}$$

Here, $\omega_0$ is the anti-symmetric mode frequency and $\gamma$ the associated damping, and $\Omega_0$ and $\Gamma$ denote the fundamental flexural mode frequency and damping respectively. $J$ is the torsion moment of inertia of the resonator, $B_\perp = B\sin(\phi)$, $\lambda_J = \frac{J}{Md^2}$, $M$ is the flexural modal mass of the resonator, $L$ is the length of the portion of outer electrode which is parallel to the central wire, and $d$ is the distance from the central wire to the outer electrode. The first term in Eq. 2 results from the transverse motion of the oscillator due to the Lorentz force exerted on the central wire, and it is proportional to $B_\perp^2$. The second term is the torsion response on resonance resulting from the applied spin-torque , and it is proportional to $B_\perp$. As shown in the supplementary material section B, these two modes enter in the induced voltage in a linear superposition with a phase difference of $\pi$, causing a dip in the response at magnetic field $B_\perp^* = \frac{P_I}{4\pi}\frac{\phi_0}{Ld}\frac{\lambda_\gamma}{\lambda_J}$. The position of the dip in the voltage amplitude is, up to geometrical pre-factors, a measure of the spin polarization. We measure the amplitude and phase of the signal $V(\omega_0)$ as a function of the applied magnetic field $B$, the driving current $I$, and the tilt angle $\phi$. For clarity, the model presented here does not include the capacitive cross-talk coupling between the drive and detection electrodes in our setup, the analysis of which results in a small correction to the detected voltage signal (supplementary section E). The amplitude of Eq. 2 is plotted in Fig. 2b, and we use this form to fit our detected voltage signal.

In addition to the main with a FM-NM (Co-Au) interface on the central wire, we have also fabricated and measured an equivalent control sample without a FM-NM interface. In the control sample, the metallic electrode on the central wire is entirely cobalt (Co). Fig. 3 shows the field dependence of the measured voltage amplitude for the Au-Co (plot a) and Co (plot b) samples at the stage tilt angle $\phi = 25^\circ$. The response from the Co sample is due only to the Lorentz force excitation and has the expected $B^2$ dependence. On the contrary, the signal from the Au-Co sample contains a contribution from the response to the spin torque at the FM-NM interface. The control sample check rules out effects such as magneto-resistance or the Wiedemann torque[24], which is expected to be very small in the present experiment.

In Fig. 4a, we plot the measured voltage amplitude for different driving currents at stage angle $\phi = 25^\circ$ for the Au-Co sample. We observe that the response varies linearly with current (Fig. 4b), as expected from Eq. 2. We also plot $V(\omega_0)/\sin^2\phi$ at various stage angles $\phi$ in Fig. 4c and 4d. This normalized response approaches the $B^2$ Lorentz form as the central wire is tilted away from the applied field ($\phi$ increases) since the polarization $P_I(B) = P_I \cos\phi$ along the wire vanishes. At low tilt angle $\phi$, where the polarization is mainly along the wire, the spin torque manifests itself through a well predicted deviation from the quadratic field dependence of the torsion response amplitude (Fig. 4c). The typical spin-flip torque detected in our nanowire carrying a current of $1\ \mu\text{A}$ is equal to $2.3\times10^{-22}$ N-m.

From numerical fitting of the data we extract the polarization parameter $P = 0.85 \pm 0.04$, where the error arises from the uncertainty in the estimates of mechanical parameters. Cobalt is known to be a strong ferromagnet with all majority spin *d*-bands filled and nearly negligible spin polarization of *sp*-electrons. Therefore, we can identify the current polarization *P* directly with the relative contribution of *d*-electrons (supplementary section A). In comparison with previous experiments[27,28], ours is a novel technique to measure *P* independently, as it does not involve superconducting contacts. In next generation experiments, the measurement of *P* will require a complete calibration protocol to reach the necessary level of precision.

We express our sensitivity in terms of the minimum detectable number of spins, where the associated oscillator displacement from a single spin flip event is $x_1 = \frac{\hbar d}{2J_{SF}\gamma}$. The smallest detected signal in our experiment corresponds to 76,000 spin-flip events ( $I = 0.75\ \mu A$ with acquisition time of 1 sec). To estimate the expected sensitivity of our device, we have performed a detailed theoretical analysis of noise (supplementary section C and D). We show that the two dominant sources of noise in our setup are the preamplifier noise (effective noise temperature $T_N = 92$ K) and the thermal noise of the mechanical mode, estimated using the classical fluctuation-dissipation theorem. The preamplifier noise and thermal noise correspond to equivalent torque noise spectral densities of $S_T^{1/2}(amp) = 5.0 \times 10^{-23}$ N-m/$\sqrt{\text{Hz}}$ and $S_T^{1/2}(th) = 3.0 \times 10^{-24}$ N-m/$\sqrt{\text{Hz}}$ respectively. The preamplifier noise determines our limiting sensitivity of 23,500 spin-flip events per $\sqrt{Hz}$. Since our experimental setup is not perfectly optimized, it exhibits a slightly higher noise than these theoretical estimates. Significant enhancement in the detector sensitivity is expected using ultra-low-noise preamplifiers in our next-generation magnetometer design with higher resonance frequencies and lower measurement temperature.

Here, we have demonstrated spin-torque detection with a sensitivity of $10^{-22}$ N-m$/\sqrt{\text{Hz}}$ in a FM-NM hybrid torsion oscillator. This level of torque sensitivity competes favorably with the $10^{-21}$ N-m$/\sqrt{\text{Hz}}$-range sensitivity in optical-tweezers approaches used for molecular torque measurements. In addition, our approach paves the way to the development of new devices combining spintronics and nanomechanics, with applications from molecular torque detection and measurement[29] to nanomechanical tests of spintronics effects[30]. Future work will require improvement in fabrication, measurement, and characterization, as well as development of calibration protocols for precise quantitative analysis.

**Acknowledgements**

This work was supported by National Science Foundation (DMR-0346707) under the NSF-EC Cooperative Activity in Materials Research. SK acknowledges support by DFG SFB668 B2 and DFG SFB508 B9. We thank Mark Johnson, Igor Zutic, Tim Wehling, John Wei, Laurent Saminadayar, and Claudio Chamon for helpful discussions.

**Author Contributions**

All authors discussed the results and commented on the manuscript.

All correspondence should be addressed to PM at mohanty@buphy.bu.edu.

[1] Richardson, O. W. A mechanical effect accompanying magnetization. *Phys. Rev.* **26**, 248 (1908).

[2] Einstein, A. & de Hass, W.J. Experimenteller Nachweis der ampere'schen molekularstroeme. *Verhandlungen der Deutschen Physikalischen Gesellschaft* **17**, 152 (1915).

[3] Beth, R. A. Mechanical detection and measurement of the angular momentum of light, *Phys. Rev.* **50**, 115 (1936).

[4] Barnett, S. J. New researches in gyromagnetism. *Phys. Rev.* **66**, 224 (1944).

[5] Kittel, C. On the gyromagnetic ratio and spectroscopic splitting factor of ferromagnetic substances. *Phys. Rev.* **76**, 743 (1949).

[6] Scott, G. G. A precise mechanical measurement of the gyromagnetic ratio of Iron. *Phys. Rev.* **82**, 542 (1952).

[7] Wallis, T. M., Moreland, J. and Kabos, P., Einstein-de Haas effect in a NiFe film deposited on a microcantilever, *Appl. Phys. Lett.* **89**, 122502 (2006).

[8] Ascoli, C., et al., Micromechanical detection of magnetic resonance by angular momentum absorption, *Appl. Phys. Lett*. **69**, 3920 (1996).

[9] Slonczewski, J.C., Current-Driven Excitation of Magnetic Multilayers, *J. Magn. Magn. Mater.* **159**, L1 (1996).

[10] Berger, L., Emission of Spin Waves by a Magnetic Multilayer Traversed by a Current, *Phys. Rev. B* **54**, 9353 (1996).

[11] Sun, J. Z. Current-driven magnetic switching in manganite trilayer junctions. *J. Magn. Mater.* **202**, 157−162 (1999).

[12] Myers, E. B., Ralph, D. C., Katine, J. A., Louie, R. N. and Buhrman, R. A. Current-induced switching of domains in magnetic multilayer devices. *Science* **285**, 867−870 (1999).

[13] Tsoi, M. et al. Generation and detection of phase-coherent current-driven magnons in magnetic multilayers. *Nature* **406**, 46−48 (2000).

[14] Wegrowe, J.-E. et al. Exchange torque and spin transfer between spin polarized current and ferromagnetic layers. *Appl. Phys. Lett.* **80**, 3775−3777 (2002).

[30] Weiss, P., Magnetic Overthrow, *Science News* **169**, 1 (2006).

**Figure 1** Spin-torsion oscillator diagram. **a** SEM micrograph of the nanomechanical device with the insert showing the FM-NM interface (colorized, Co – red, Au - yellow). The oscillator overall dimensions are 12μm x 6 μm with 500 nm thickness. The central wire is 300 nm wide, with 50 nm thick Au deposited on the non-magnetic side and 50-nm thick Co on the ferromagnetic side. The fully suspended structure is clamped rigidly at the large support pads and placed in $10^{-6}$ Torr vacuum of a dilution refrigerator at 110 mK. **b** Finite element simulation of the anti-symmetric torsion mode showing the color-coded elastic strain localized on the central wire. For nearly equal rotational moments of inertia of the inner and outer torsion elements, the strain in the anti-symmetric mode provides optimum amplitude coupling and minimized dissipation. The measured resonance signal is a Lorentzian peak centered at $f_0$ = 6.71 MHz with $Q$ = 28000. **c** Measurement set-up diagram showing the tilting of the sample with respect to the axial magnetic field $B$. The sample stage can be controllably tilted through an angle $\phi$ from 0 to 90 degrees. The parallel component of the magnetic field $B_{\parallel}$ polarizes the cobalt nanowire, while the perpendicular component induces the detected signal $V_{emf}$ on the outer electrode. We drive the central wire on resonance and detect the induced signal on the outer electrode using a lock-in amplifier.

**Figure 2** Diagram and theoretical modeling of the spin-torsion mechanism. **a** At the interface of a ferromagnetic (FM) and non-magnetic (NM) segments of a quasi-1D nanowire, the spin-polarized charge carriers undergo spin-relaxation. An external magnetic field $B$ is applied to magnetize the ferromagnet along the easy axis. The change in the spin direction of the electrons produces mechanical torque on the crystal lattice as a result of angular momentum conservation. The mechanical torque is directed along the wire. **b** Analytical form of the expected voltage signal (amplitude of Eq. 2 in the text). The signature of the spin-torque is the deviation of the response from the $B^2$ Lorentz form.

**Figure 3** The measured response of two equivalent devices: one with and one without a FM-NM interface. Magnetic field sweeps were performed on resonance at stage angle $\phi$ = 25 degrees and current $I = 5.5\ \mu A$. **a** Measurement of the Au-Co sample shows excellent agreement with the analytical fit to the spin-torque response amplitude in Eq. 2 (red curve). The $B^2$ Lorentz response background (blue curve) is included for reference. **b** Measurement of the Co control sample, containing no FM-NM interface, follows closely the expected Lorentzian response (red line fit), showing quadratic *B*-dependence.

**Figure 4** The current and stage-angle dependence of the voltage response in the Au-Co sample. **a** We plot the resonator response at various driving currents at a fixed stage angle of $\phi$ = 25 degrees. The numerical fits (amplitude of Eq. 2) are shown in black. **b** We verify the linear current dependence of the spin torque response at two values of the magnetic field and tilt angle $\phi$ = 25 degrees. **c** We normalize the Lorentz voltage response by $\sin^2\phi$ to show the variation with $\phi$, and plot the response at three values of the sample stage tilt. At the low tilt angle $\phi = 25$ degrees, the applied field polarizes the ferromagnet along the wire, and the resulting spin-torque is manifested as increased torsion response in the low field range up to 2 Tesla. Above this field range the field dependent phase of the response in Eq. 2 causes the amplitude to drop below the $B^2$ Lorentz background. The signature of the spin torque decreases rapidly with tilt angle as the magnetization along the central wire axis decreases and the transverse field $B_\perp$ grows, moving the predicted dip in the voltage to smaller fields. **d** We show the variation of the normalized resonator response with stage angle $\phi$ for two different driving currents, showing the vanishing of the spin-torque contribution to the signal at high tilt angles as the polarizing field $B_\parallel$ vanishes. The curves are guides to the eye.

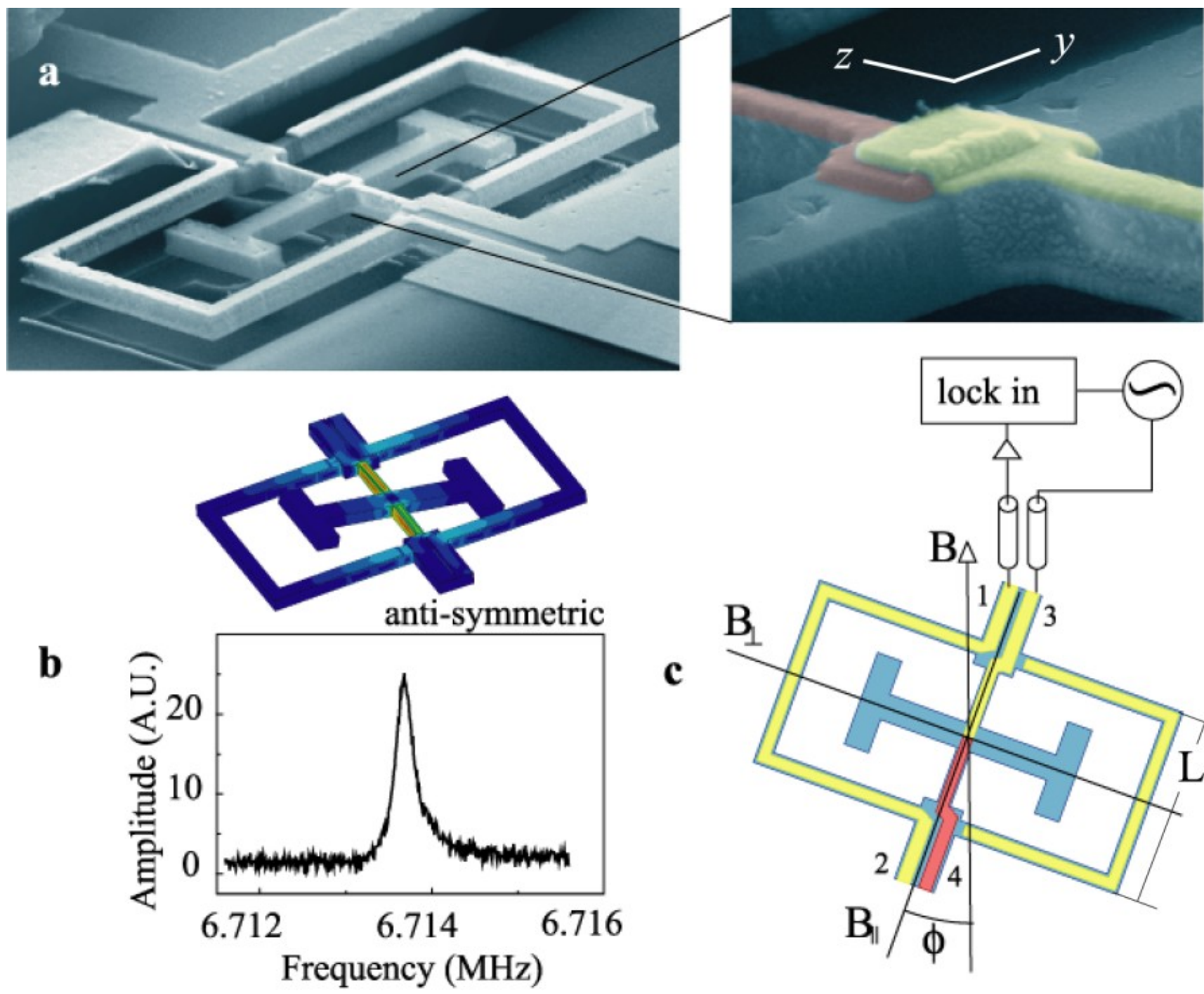
a
z
y
anti-symmetric
b
Amplitude (A.U.)
20
10
0
6.712
6.714
6.716
Frequency (MHz)
c
lock in
B
B⊥
B∥
1
2
3
4
L
ϕ

**Figure 1**

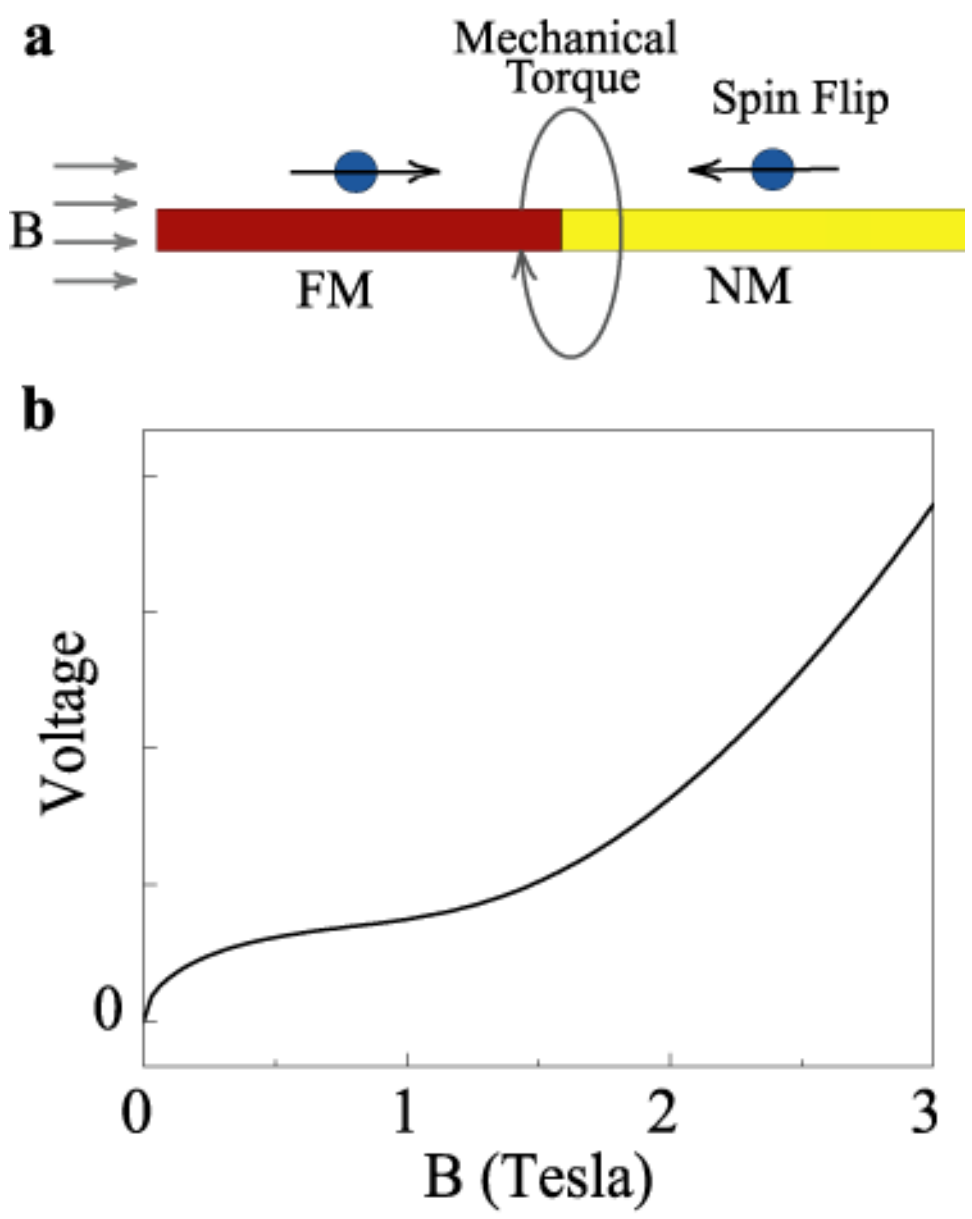

**Figure 2**

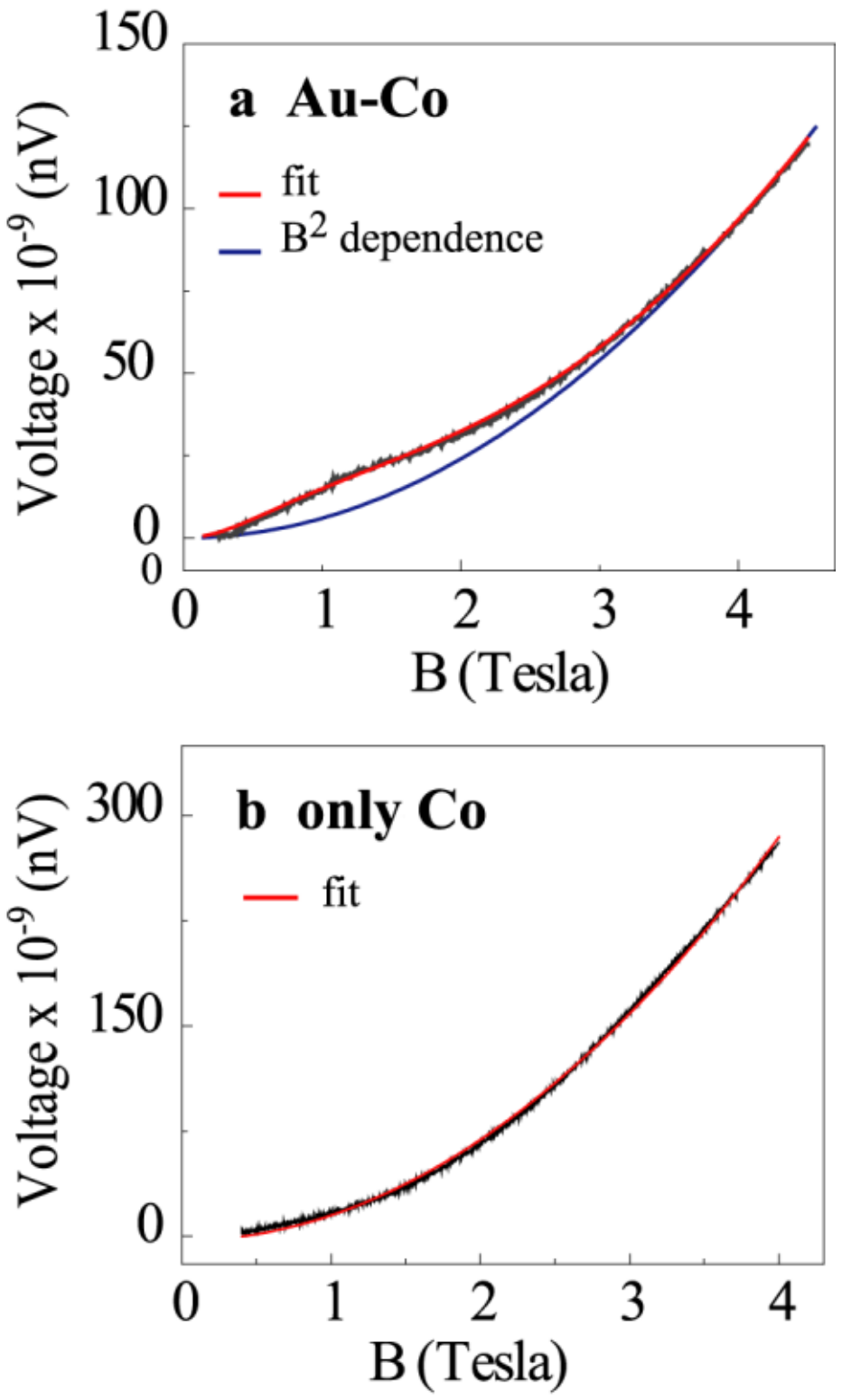

a Au-Co
fit
B2 dependence
Voltage x 10-9 (nV)
150
100
50
0
0
0
1
2
3
4
B (Tesla)
b only Co
fit
Voltage x 10-9 (nV)
300
150
0
0
1
2
3
4
B (Tesla)

**Figure 3**

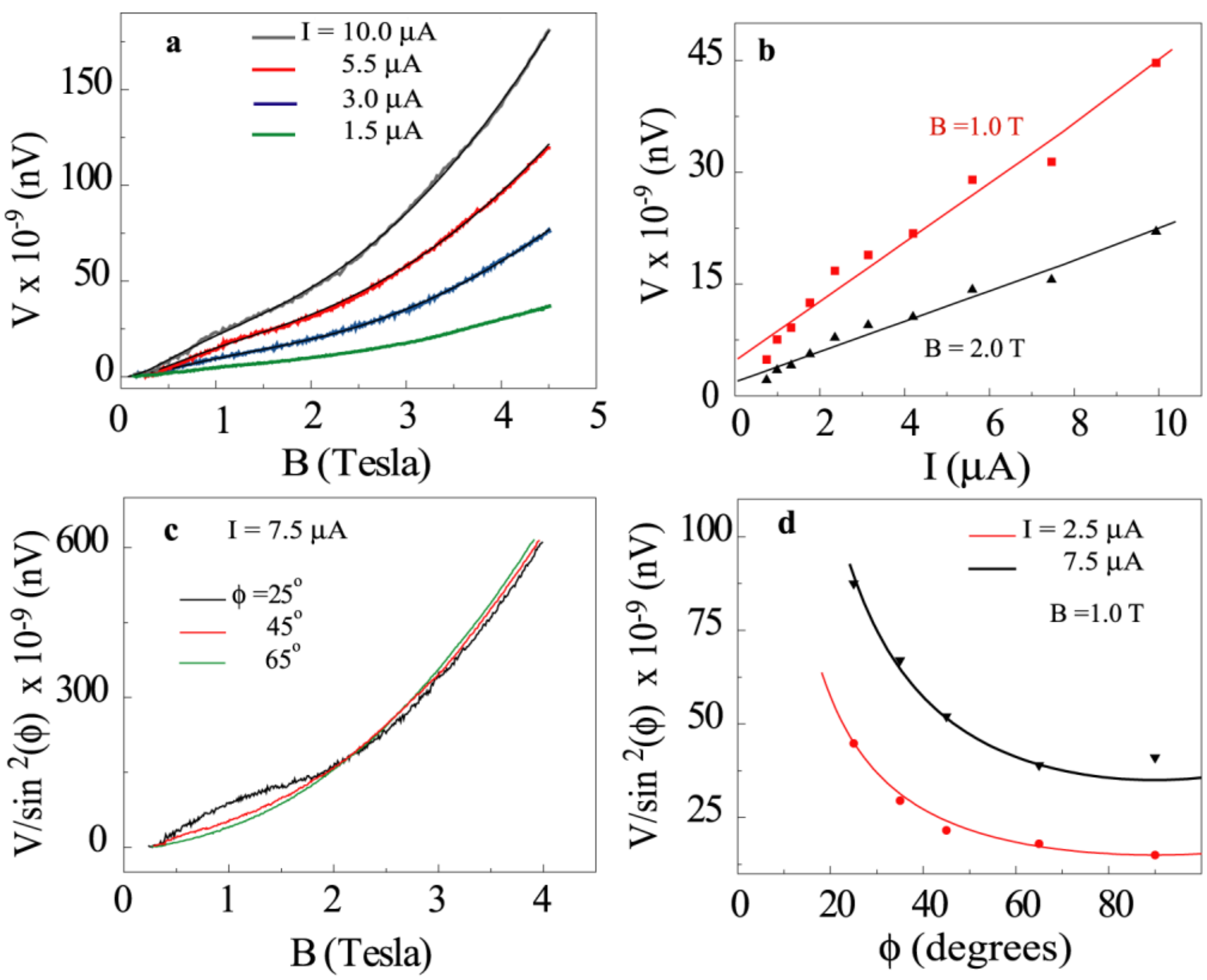

a
I = 10.0 μA
5.5 μA
3.0 μA
1.5 μA
V x 10^-9 (nV)
B (Tesla)
b
B =1.0 T
B = 2.0 T
V x 10^-9 (nV)
I (μA)
c
I = 7.5 μA
ϕ =25°
45°
65°
V/sin^2(ϕ) x 10^-9 (nV)
B (Tesla)
d
I = 2.5 μA
7.5 μA
B =1.0 T
V/sin^2(ϕ) x 10^-9 (nV)
ϕ (degrees)

**Figure 4**